\documentclass[prd,aps,eqsecnum,floatfix,nofootinbib,preprint,tightenlines]{revtex4}

\usepackage{latexsym}
\usepackage{graphicx}
\usepackage{color}
\usepackage{amsmath}
\usepackage{hyperref}

\def\bibi{\bibitem}

\let\br=\u                      

\let\inodot=\i

\def\a{\alpha}
\def\b{\beta}

\def\d{\delta}
\def\g{\gamma}
\def\h{\eta}
\def\i{\iota}

\def\m{\mu}
\def\n{\nu}

\def\p{\pi}                     
\def\r{\rho}                    
\def\s{\sigma}                  
\def\t{\tau}

\def\D{\Delta}

\def\P{\Pi}

\def\cbo{{\,\raise-.15ex\Sc [\,}}                       
\def\gtap{\raisebox{-.4ex}{\rlap{$\sim$}} \raisebox{.4ex}{$>$}}   

\def\ddt#1{{\buildrel {\hbox{\LARGE .\kern-2pt.}} \over {#1}}}

\def\ie{\mbox{\it i.e.}}

\def\half{{1\over 2}}

\long\def\symbolfootnote[#1]#2{\begingroup%
\def\thefootnote{\fnsymbol{footnote}}\footnote[#1]{#2}\endgroup}

\long \def \blockcomment #1\endcomment{}

\def\bP{{\overline{\P}}}
\def\br{{\overline{\r}}}
\def\hatw{{\hat{w}}}

\begin{document}

\thispagestyle{empty}

\begin{center}
\vspace*{5mm}
\begin{boldmath}
{\large\bf Low-energy constants and condensates from ALEPH hadronic $\tau$ 
decay data}
\end{boldmath}
\\[10mm]
Diogo Boito,$^a$ Anthony Francis,$^b$
Maarten Golterman,$^c$ Renwick Hudspith,$^b$ Randy Lewis,$^b$
Kim Maltman,$^{d,e}$ Santiago Peris$^f$
\\[8mm]
{\small\it
$^a$Instituto de F\'\inodot sica, Universidade de S\~ao Paulo,
Rua do Mat\~ao Travessa R, 187, 05508-090,\\ S\~ao Paulo, SP, Brazil
\\[5mm]$^b$Department of Physics and Astronomy,
York University\\  Toronto, ON Canada M3J~1P3
\\[5mm]
$^c$Department of Physics and Astronomy,
San Francisco State University\\ San Francisco, CA 94132, USA
\\[5mm]
$^d$Department of Mathematics and Statistics,
York University\\  Toronto, ON Canada M3J~1P3
\\[5mm]
$^e$CSSM, University of Adelaide, Adelaide, SA~5005 Australia
\\[5mm]
$^f$Department of Physics, Universitat Aut\`onoma de Barcelona\\ E-08193 Bellaterra, Barcelona, Spain}
\\[10mm]
{ABSTRACT}
\\[2mm]
\end{center}

\begin{quotation}
We determine the NLO chiral low-energy constant $L_{10}^r$, and 
combinations $C_{12}^r\pm C_{61}^r+C_{80}^r$, $C_{13}^r-C_{62}^r+C_{81}^r$,
$C_{61}^r$, and $C_{87}^r$, of the NNLO chiral low-energy constants
incorporating recently revised ALEPH results for the non-strange vector 
($V$) and axial-vector ($A$) hadronic $\tau$ decay distributions and recently
updated RBC/UKQCD lattice data for the non-strange $V-A$ two-point function. 
In the $\overline{\rm MS}$ scheme, at renormalization scale $\m=770\ \mbox{MeV}$, 
we find $L_{10}^r=-0.00350(17)$, 
$C_{12}^r+C_{61}^r+C_{80}^r=0.00237(16)\ \mbox{GeV}^{-2}$, 
$C_{12}^r-C_{61}^r+C_{80}^r=-0.00056(15)\ \mbox{GeV}^{-2}$, 
$C_{13}^r-C_{62}^r+C_{81}^r=0.00046(9)\ \mbox{GeV}^{-2}$, 
$C_{61}^r=0.00146(15)\ \mbox{GeV}^{-2}$, and
$C_{87}^r=0.00510(22)\ \mbox{GeV}^{-2}$.
With errors here at or below the level expected for contributions 
of yet higher order in the chiral expansion, the analysis exhausts the
possibilities of what can be meaningfully achieved in an NNLO analysis.
We also consider the dimension six and dimension eight coefficients in the
operator product expansion in the $V-A$ channel.
\end{quotation}

\newpage
\section{\label{introduction} Introduction}
Recently, a revised version \cite{ALEPH13} of the ALEPH data \cite{ALEPH98}
for the non-strange vector ($V$) and axial vector ($A$) spectral distributions 
obtained from measurements of hadronic $\tau$ decays
became available. These results corrected a problem, uncovered 
in Ref.~\cite{TAU10}, in the publicly posted 2005 and 2008 versions 
of the correlations between different energy bins.\footnote{The 
corrected data may be found 
at http://aleph.web.lal.in2p3.fr/tau/specfun13.html.}
In Ref.~\cite{alphas14} we analyzed these data in order to extract the strong
coupling at the $\t$ mass, $\a_s(m_\t^2)$, as well as OPE condensates,
following the strategy previously developed in Refs.~\cite{US1,US2}. 
This analysis leads to a complete description of the $V$ and $A$ 
spectral functions as a function of the energy-squared $s$, including 
the region $s>m_\t^2$.

Complete knowledge of the spectral functions allows one to construct 
a representation of the (subtracted) vacuum polarizations in both 
channels, and the unsubtracted vacuum polarization in the $V-A$ channel, 
as a function of $s$. Combining 
these representations 
with the analytic expressions derived from chiral perturbation theory 
(ChPT), which are known to next-to-next-to-leading order (NNLO) \cite{ABT} 
allows us to extract the low-energy constant (LEC) $C_{87}$, 
and a linear combination of $L_{10}$, $C_{12}-C_{61}+C_{80}$ and 
$C_{13}-C_{62}+C_{81}$ from conventional chiral sum rules for the 
non-strange $V-A$ channel \cite{DGHS,GPP}, and $C_{12}+C_{61}+C_{80}$ and
a linear combination of $L_{10}$ and $C_{12}-C_{61}+C_{80}$ from 
flavor-breaking ($ud-us$) chiral sum rules \cite{DK} in the $V\pm A$ 
channels. These determinations employ as input existing values of $L_5$ 
and $L_9$, estimates existing from both 
phenomenology \cite{bj11,bt02} and the lattice \cite{FLAG,MILC9A}.
In order to disentangle further $L_{10}$, $C_{12}-C_{61}+C_{80}$ and 
$C_{13}-C_{62}+C_{81}$, we exploit the dependence of the coefficients 
of these LEC combinations appearing in the $V-A$ 
polarization on the pion and kaon masses using data for this polarization 
from the lattice \cite{bgjmp13,Boyleetal14}.\footnote{We
hope to revisit the determination of $L_5$ and $L_9$ from the 
$ud$ and $us$ $V$ polarizations 
in the future, but it remains to be seen whether the errors from such 
an analysis are competitive with those of Refs.~\cite{bj11,bt02,FLAG,MILC9A}.   
Here we will use values from the literature.}

The goal of this article is to update the analysis of 
Refs.~\cite{bgjmp13,Boyleetal14,GMPNNLO}, replacing the 
experimental data for the non-strange spectral functions, which previously 
came from OPAL \cite{OPAL}, with the revised data from Ref.~\cite{ALEPH13}, and, 
at the same time, updating the lattice input of Ref.~\cite{Boyleetal14}.
The expectation is that the errors
on $L_{10}$ and the accessible NNLO combinations will decrease, because the
ALEPH data are more precise, especially in the low-energy region. 
Improvements in the lattice data should also help in
the process of disentangling $L_{10}$ from the combinations
$C_{12}-C_{61}+C_{80}$ and $C_{13}-C_{62}+C_{81}$, described previously in
Ref.~\cite{bgjmp13}. In the current analysis, we employ a slightly 
different input for $L_5$, choosing now to use the $2+1$-flavor estimate 
of Ref.~\cite{MILC9A}, $L_5^r(\m=770~\mbox{MeV})=0.84(38)\times 10^{-3}$.
This value is the $2+1$-flavor estimate adopted in Ref.~\cite{FLAG}, and straddles 
several nominally more precise, but not mutually consistent, estimates, 
including the result $L_5^r(\m=770~\mbox{MeV})=0.58(13)\times 10^{-3}$ 
of Ref.~\cite{bj11} used in our previous analysis.

While the main emphasis here is on the LECs of ChPT,
we will also update our estimates for the operator product expansion (OPE)
condensates $C_{6,V-A}$ and $C_{8,V-A}$, which are order parameters
(in contrast to the analogous condensates in the $V+A$ channel).   
By their nature, these condensates are sensitive to the high-energy part 
of the spectral function, which is less well known. Therefore, it is 
interesting to compare the values for the condensates
obtained from the OPAL and ALEPH data.

\section{\label{theory} Theory compendium}
In this section, we briefly collect the definitions and relations between 
quantities needed in order to present our results. More detailed overviews 
can be found in Refs.~\cite{bgjmp13,GMPNNLO} and references therein. We first 
consider the required sum-rule results, and then collect relevant results 
from ChPT.

\subsection{\label{sum rules} Sum rules}
The vacuum polarizations, $\Pi^{(J)}_T(Q^2)$, with $T=V,\, V\pm A$ 
and $J=0,1$, are the spin $J$ scalar parts of the corresponding 
current-current two-point functions, $\Pi^T_{\mu \nu}(q)$, and are related 
to the corresponding spectral functions, $\rho^{(J)}_T(s)$, by finite-energy
sum rules (FESRs). The flavor $ud$ and $us$ versions of these spectral 
functions can be obtained from experimental differential hadronic $\tau$ decay 
distribution data for energies up to the $\tau$ mass.
Above the $\t$ mass one needs a theoretical representation, and we 
will use the one obtained in Ref.~\cite{alphas14}. 

First, let us consider the non-strange $V-A$ channel, for which the 
vacuum polarization obeys the unsubtracted dispersion relation
\begin{eqnarray}
\label{defPiLR}
\P_{V-A}(Q^2)&\equiv&\P^{(0+1)}_{V-A}(Q^2)=
\int_0^\infty ds\;\frac{\r_V(s)-\r_A(s)}{s+Q^2}\ ,\\
\P^{(0+1)}_{V-A}(Q^2)&\equiv&
-\frac{1}{3q^2}\left(g^{\m\n}-\frac{4q^\m q^\n}{q^2}\right)
\P^{V-A}_{\m\n}(q)\ ,\nonumber
\end{eqnarray}
where the Euclidean momentum-squared $Q^2=-q^2$.
Here $\r_V$ ($\r_A$) is the
non-strange, $I=1$ vector (axial) spectral function summing the 
angular momentum $J=1$ and $J=0$ contributions. Generalizing the 
definition of $\P_{V-A}(Q^2)$, we also define functions
$\P_{V-A}^{(w)}$, to be used in the restricted 
sense employed below, involving
additional polynomial weight factors $w(y)$:
\begin{equation}
\label{defPiLRw}
\P_{V-A}^{(w)}(Q^2)=\int_0^\infty ds\;w(s/s_0)\;\frac{\r_V(s)-\r_A(s)}{s+Q^2}
\end{equation}
with $0<s_0\le m_\t^2$. In what follows, we will use the 
weights
\begin{equation}
\label{weights}
w_k(y)=(1-y)^k\ ,\quad k=1,2\ .
\end{equation}
In evaluating the integral in Eq.~(\ref{defPiLRw}), we will use the ALEPH
experimental data for the spectral functions for $s\le s_0$, and the 
duality-violating (DV) part
\begin{equation}
\label{approx}
\r_V(s)-\r_A(s)\approx\r^{\rm DV}_V(s)-\r^{\rm DV}_A(s)\ ,\qquad s\ge s_0\ ,
\end{equation}
above $s_0$, with $\r^{\rm DV}_V(s)-\r^{\rm DV}_A(s)$ equal to $1/\p$ times the imaginary part of $\P_{V-A}^{\rm DV}(Q^2)$, defined, in turn, from
\begin{equation}
\label{split}
\P_{V-A}(Q^2)=\P_{V-A}^{\rm OPE}(Q^2)+\P_{V-A}^{\rm DV}(Q^2)\ ,
\end{equation}
for $\vert Q^2\vert \ge s_0$, where  the OPE part has the form
\begin{equation}
\label{highQ2}
\P_{V-A}^{\rm OPE}(Q^2)=\sum_{k=1}^\infty\;\frac{C_{2k,V/A}}{(Q^2)^k}\ ,
\end{equation}
with $C_{2k,V/A}$ the OPE coefficients. We will always assume that we
can choose $s_0$ smaller than $m_\t^2$, but large enough that the 
separation~(\ref{split}) into OPE and DV parts makes sense.   
We use a model for the DV parts of
the spectral functions:
\begin{equation}
\label{ansatz}
\r_{V/A}(s)=e^{-\d-\g s}\sin{(\a+\b s)} ,
\end{equation}
with $\a$, $\b$, $\g$, and $\d$ parameters which differ
in the $V$ and $A$ channels. The form of this {\it ansatz} was motivated 
in Ref.~\cite{CGPDV}, and it was shown in Refs.~\cite{alphas14,US1} that this 
model can be used to successfully parametrize the resonance structure
in the data for $s\,\gtap\, 1.4$~GeV$^2$. Here we will fix the values
of the $V$ and $A$ DV parameters using the results of the
FOPT fit of Table~IV of Ref.~\cite{alphas14} with 
$s_{\rm min}=1.55$~GeV$^2$.\footnote{Essentially identical 
results are obtained if one employs instead the results of the 
corresponding CIPT fit.}

In what follows we will denote by $\bP_A$, $\bP_A^{(w)}$ 
and $\br_A$ the versions of $\P_A$, $\P_A^{(w)}$ and $\r_A$
from which the pion pole contribution has been subtracted. Analogous
pion-pole-subtracted versions of the $V\pm A$ polarizations and spectral
functions are denoted $\bP_{V\pm A}=\P_V\pm\bP_A$, 
$\bP_{V\pm A}^{(w)}=\P_V^{(w)}\pm\bP_A^{(w)}$ and $\br_{V\pm A}=\r_V
\pm\br_A$.

For $Q^2<4 m_\pi^2$, $\bP_{V-A}(Q^2)$ admits a Taylor expansion,
and we can thus define the intercept and slope at $Q^2=0$,
\begin{equation}
\label{lowQ2}
\bP_{V-A}(Q^2)=-8L_{10}^{\rm eff}-16C_{87}^{\rm eff}Q^2+O(Q^4)\ .
\end{equation}
Employing FESRs for the weights~(\ref{weights}) and analytic results for
the OPE coefficients $C_{2,V-A}$ and $C_{4,V-A}$ \cite{FNR,NLOC4}, 
it follows that \cite{US1}
\begin{eqnarray}
\label{alt}
&&\hspace{-0.4cm}-8L_{10}^{\rm eff}=\bP_{V-A}(0)=
\bP_{V-A}^{(w_1)}(0)+\frac{2f_\p^2}{s_0}\\
&&\hspace{0.2cm}=\bP_{V-A}^{(w_2)}(0)+
\frac{4f_\p^2}{s_0}\left[1-\frac{17}{16\p^2}\left(\frac{\a_s(s_0)}{\p}\right)^2\frac{m_u(s_0)m_d(s_0)}{f_\p^2}
-\frac{m_\p^2}{2s_0}\left(1+\frac{4}{3}\frac{\a_s(s_0)}{\p}\right)
\right]\ .
\nonumber
\end{eqnarray}
Since the terms proportional to $\a_s(s_0)$ lead to effects smaller than the
experimental errors, we will omit these terms from the actual analysis leading
to our results for $L_{10}^{\rm eff}$ and $C_{87}^{\rm eff}$.

In addition to the information obtained from the flavor 
$ud$ V-A channel, further constraints can be obtained
from inverse-moment FESRs (IMFESRs) for the flavor-breaking differences
\begin{equation}
\label{Pidiff}
\D\P_T(Q^2)\equiv\P^{(0+1)}_{ud;T}(Q^2)-\P^{(0+1)}_{us;T}(Q^2)\ ,
\end{equation}
defined in Ref.~\cite{GMPNNLO}. Note that the $\Delta \Pi_T(Q^2)$ 
are finite and satisfy unsubtracted dispersion relations.
The IMFESRs of Ref. [17] have the forms
\begin{eqnarray}
\label{vpmaimsralt}
\D\P_V(0) &=&
\int_{s_{\rm th}}^{s_0}ds\, {\frac{w(s/s_0)}{s}}\,\Delta{\rho}_V(s)
+{\frac{1}{2\pi i}}\,\oint_{\vert z\vert = s_0} dz\,
{\frac{w(z/s_0)}{z}}\, \left[\Delta\Pi_V (-z)\right]^{\rm OPE}\ ,\label{dkvsr}
\nonumber
\\
\D\bP_{V\pm A}(0) &=& 
\int_{s_{\rm th}}^{s_0}ds\, {\frac{w(s/s_0)}{s}}\,\Delta\overline{\rho}_{V\pm A}(s)
\pm \left[{\frac{f_K^2}{s_0}}\, f^w_{\rm res}(y_K)\,
-\, {\frac{f_\pi^2}{s_0}}\, f^w_{\rm res}(y_\pi )\right]
\nonumber\\
&&\ \ \ \ \ +\,{\frac{1}{2\pi i}}\,\oint_{\vert z\vert = s_0} dz\,
{\frac{w(z/s_0)}{z}}\, \left[\Delta\Pi_{V\pm A} (-z)\right]^{\rm OPE}\ ,
\end{eqnarray}
in which $s_{\rm th}$ is the continuum threshold $4m_\p^2$, 
$y_\p=m_\p^2/s_0$, $y_K=m_K^2/s_0$, and
\begin{equation}
\label{fdef}
f(y)=\frac{2}{y}\left(w(0)-w(y)\right)\ .
\end{equation}
As long as we retain the exact $\D\P(-z)$ in the contour 
integrals of Eq.~(\ref{vpmaimsralt}), the full 
right-hand sides are necessarily independent of the weight choice $w$,
provided we restrict our attention to $w(y)$ all having $w(0)=1$.
In Eq.~(\ref{vpmaimsralt}) we dropped the DV term from the split~(\ref{split}), 
and kept only the OPE contribution. It is reasonable to do so, because the
only weights we will use will be triply pinched, \ie, contain a factor 
$(z-s_0)^3$, which suppresses DVs strongly.
In our analysis, we will use the weights
\begin{eqnarray}
\label{IMweights}
\hatw(y)&=&(1-y)^3\ ,\\
w_{\rm DK}(y)&=&(1-y)^3\left(1+y+\half y^2\right)
=1-2y+\half\left(y^2+y^3+y^4-y^5\right)\ .\nonumber
\end{eqnarray}
The IMFESR with $w_{\rm DK}$ was first considered in Ref.~\cite{DK}.

The quantities $\bP_{V-A}(0)$, $\bP'_{V-A}(0)$, $\D\P_V(0)$ and 
$\D\bP_{V\pm A}(0)$ can be obtained from hadronic $\t$-decay data,
which yield the spectral functions $\r_{V/A}(s)$, either directly, 
for $s<m_\t^2$, or, where it is needed for $s>m_\t^2$ in the $V-A$
channel analysis, indirectly through Eq.~(\ref{approx}), 
using the values of the DV parameters obtained from the fits to 
these same data,
described in Ref.~\cite{alphas14}. In addition, one needs the experimental values
of $m_\p$, $f_\p$,\footnote{Our normalization is such that 
$f_\p=92.21(14)$~MeV.}
$m_K$ and $f_K$. For a detailed discussion of the OPE contributions to
Eq.~(\ref{vpmaimsralt}), we refer to Ref.~\cite{GMPNNLO}.\footnote{See, in particular,
Sec.~III.B of that reference.} A key point is that the numerical contributions
from the OPE terms to $\D\P_V(0)$ and $\D\bP_{V\pm A}(0)$ are very small.
That implies that even if these OPE contributions are not very well known,
and one has therefore to include very conservative estimates 
of their errors in the total error budget, the impact 
on our final errors is minor.

As already noted above, we have dropped DV contributions in 
Eq.~(\ref{vpmaimsralt}). The reason is that these are very suppressed for the 
weights~(\ref{IMweights}), which are triply pinched, and moreover
suppress the large-$s$ region by an additional factor $1/s$.   
Since the $s_0$ dependence from both the OPE and DVs is non-trivial,
our treatment of the OPE and the omission of DVs can be tested by
checking the $s_0$ independence of $\D\P_V(0)$, $\D\bP_{V+A}(0)$ and
$\D\bP_{V-A}(0)$ produced using the right-hand side of the 
corresponding IMFESR.

\subsection{\label{chpt} Chiral perturbation theory}
The motivation for considering the quantities $L_{10}^{\rm eff}$, 
$C_{87}^{\rm eff}$, $\D\P_V(0)$, and $\D\bP_{V\pm A}(0)$ is that they 
all depend on NLO and NNLO LECs of the chiral effective theory, and 
thus yield information on the QCD values for these LECs if sufficiently 
accurate data (from experiment or lattice QCD) are
available.   Here we will collect the relevant NNLO ChPT expressions needed
in order to connect the quantities defined in the previous subsection to the
LECs of ChPT.   

The representation of $\bP_{V-A}(Q^2)$ to NNLO in ChPT has the form
\begin{eqnarray}
\label{Pifit}
\bP_{V-A}(Q^2)&=&-8\Bigl(1-4(2\m_\p+\m_K)\Bigr)L_{10}^r
+32m_\p^2\left(C_{12}^r-C_{61}^r+C_{80}^r\right)\\
&&+32(2m_K^2+m_\p^2)\left(C_{13}^r-C_{62}^r+C_{81}^r\right)
-16C_{87}^rQ^2+R_{\p K}(\m,Q^2;L_9^r)\ ,\nonumber
\end{eqnarray}
where the explicit expression for $R_{\p K}(\m,Q^2;L_9^r)$ can be 
reconstituted from the results of Ref.~\cite{ABT}.\footnote{Since the value 
of $L_9^r$ is well known \cite{bt02}, we treat the loop contribution 
proportional to this LEC as known, and we thus include
this contribution in $R_{\p K}(\m,Q^2;L_9^r)$.}
The subscript $\p K$ indicates that $R$ depends
also on $m_\p$, $m_K$ and $f_\p$, in addition to the explicitly shown 
arguments.%
\footnote{In Ref.~\cite{bgjmp13} we denoted this term simply as $R(Q^2;L_9^r)$.}
$L_{10}^{\rm eff}$ and 
$C_{87}^{\rm eff}$ are then given by:
\begin{subequations}
\label{chptconn}
\begin{eqnarray}
\bP_{V-A}(0)&=&-8L_{10}^{\rm eff}\nonumber\\
&=&-8L_{10}^r\Bigl(1-4(2\m_\p+\m_K)\Bigr)
+16(2\m_\p+\m_K)L_9^r
\label{chptconna}\\
&&+32m_\p^2\left(C_{12}^r-C_{61}^r+C_{80}^r\right)+32(2m_K^2+m_\p^2)
\left(C_{13}^r-C_{62}^r+C_{81}^r\right)\nonumber\\
&&+{\hat R}_{\p K}(\m,0)
\ ,\nonumber
\\
\bP_{V-A}'(0)&=&-16C_{87}^{\rm eff}\label{chptconnb}\\
&=&-16C_{87}^r+\frac{1}{4\p^2 f_\p^2}\left(1-\log\frac{\m^2}{m_\p^2}+
\frac{1}{3}\log\frac{m_K^2}{m_\p^2}\right)L_9^r
\nonumber\\
&&+\frac{\partial {\hat R}_{\p K}(\m,Q^2)}{\partial Q^2}\Biggr|_{Q^2=0}
\ ,\nonumber\\
\m_P&=&\frac{m_P^2}{32\p^2f_\p^2}\log\frac{m_P^2}{\m^2}\ ,\label{chptconnc}
\end{eqnarray}
\end{subequations}
where $\hat R_{\p K}(\m,Q^2)$ is the part of $R_{\p K}(\m,Q^2;L_9^r)$
independent of $L_9^r$.
Here the superscript $r$ denotes the values of LECs renormalized in the $\overline{\rm MS}$ scheme at scale
$\m$, which we will take to be $\m=770$~MeV in what follows.   

We will also need the NNLO ChPT expressions for $\D\P_T(0)$, $T=V$,
$V\pm A$, but only at physical values of $m_\p$, $m_K$ and $f_\p$.
We therefore give the expressions in terms of the LECs with the numerical
values of the coefficients for the NLO LECs $L_{5,9,10}^r$ evaluated at 
$m_\p=139.57$~MeV, $M_K=495.65$~MeV,\footnote{This is the average of the
charged and neutral kaon masses. Taking just the charged or neutral value
has no impact on our final results.}
$f_\p=92.21$~MeV, and $\m=770$~MeV:
\begin{eqnarray}
\D\Pi_V(0)&=&0.00775-0.7218\, L_5^r\,
+\, 1.423\, L_9^r\, +\, 1.062\, L_{10}^r 
\, +\, 32(m_K^2-m_\pi^2)C_{61}^r\ ,\nonumber\\
\D\bP_{V+A}(0)&=&
0.00880 -0.7218\, L_5^r\, +\, 1.423\, L_9^r
\, +\, 32(m_K^2-m_\pi^2)\left[ C_{12}^r+C_{61}^r
+C_{80}^r\right]\ ,\nonumber\\
\D\bP_{V-A}(0)&=&
0.00670-0.7218\, L_5^r\, +\, 1.423\, L_9^r\, +\, 2.125\,
L_{10}^r\nonumber\\
&&\qquad\qquad\qquad\qquad\qquad -\, 32(m_K^2-m_\pi^2)\left[ C_{12}^r-C_{61}^r
+C_{80}^r\right]\ .
\label{nlonnlocontributions}
\end{eqnarray}

Equations~(\ref{Pifit}),~(\ref{chptconn}) and~(\ref{nlonnlocontributions}) 
give access to combinations involving
the LECs $L_{10}^r$, $C_{87}^r$, and the linear combinations
\begin{eqnarray}
{\cal C}_0^r&\equiv&
32 m_\pi^2 \left( C^r_{12}-C^r_{61}+C^r_{80}\right)\ ,\nonumber\\
{\cal C}_1^r&\equiv& 32 \left( m_\pi^2+2m_K^2\right)
\left( C^r_{13}-C^r_{62}+C^r_{81}\right)\ .
\label{c0c1defns}
\end{eqnarray}
In addition, the LECs $L_5^r$, $C_{61}^r$ and the linear combination
$C^r_{12}+C^r_{61}+C^r_{80}$
appear in Eq.~(\ref{nlonnlocontributions}).
Our goal is to use the ALEPH hadronic $\t$-decay data to extract $L_{10}^r$,
$C_{61}^r$, $C_{87}^r$ and $C_{12}^r\pm C_{61}^r+C_{80}^r$, using the known values of 
$L_5^r$ and $L_9^r$. However, $L_{10}^r$ appears in a linear combination 
with ${\cal C}_0^r$ and ${\cal C}_1^r$ that does not depend on $Q^2$, and 
we thus need other information in order to disentangle $L_{10}^r$, 
${\cal C}_0^r$ and ${\cal C}_1^r$ from each other. This can be done by 
using lattice data for different values of the meson masses, and such
data are available for $T=V-A$ \cite{Boyleetal14}.\footnote{No such lattice
analysis has been performed yet for the $us$ channel, which is why we 
will use known values for $L_5^r$ and $L_9^r$ here.}

We thus will consider, following Ref.~\cite{bgjmp13},
\begin{equation}
\label{LPiVA}
\D\bP_{V-A}^{\rm L,E}(Q^2)\equiv 
\bP^{\rm L,E}_{V-A}(Q^2)-\bP_{V-A}(Q^2)\ ,
\end{equation}
the difference between the non-strange pion-pole-subtracted
$V-A$ correlator $\bP^{\rm L,E}_{V-A}(Q^2)$ evaluated on 
a lattice ensemble $E$ with $\pi$ and $K$ masses and decay constants 
different from the physical ones, and the same correlator, 
$\bP_{V-A}(Q^2)$, evaluated at the same $Q^2$, for the physical 
quark mass case. The latter is obtained from the dispersive 
representation using spectral functions obtained from the 
hadronic $\tau$ decay data.
It then follows that in terms of LECs this difference can be expressed
as
\begin{equation}
\D\bP_{V-A}^{\rm L,E}(Q^2) = \D R^{\rm L,E}_{\pi K}(\m ,Q^2; L_9^r)
\, +\, \d^{\rm L,E}_{10}\, L_{10}^r\, +\, 
\d^{\rm L,E}_0 {\cal C}_0^r\, +\, \d^{\rm L,E}_1 {\cal C}_1^r\ ,
\label{deltadeltapibar}
\end{equation}
where $\D R^{\rm L,E}(\m ,Q^2; L_9^r)$ and the $Q^2$-independent coefficients 
$\d^{\rm L,E}_{10,0,1}$ are known in terms of the lattice and physical 
meson masses and decay constants, and the chiral renormalization scale $\m$.
Evaluating Eq.~(\ref{LPiVA}) using lattice and dispersive
results, Eq.~(\ref{deltadeltapibar}) yields a constraint on $L_{10}^r$,
${\cal C}^r_0$ and ${\cal C}^r_1$ for each ensemble, $E$, and each $Q^2$.
As explained in more detail below, different choices of $Q^2$ for fixed 
$E$ provide self-consistency checks on the use of the lattice data.

\section{\label{input} Input data}
We will evaluate $\P_{V-A}(Q^2)$
and $\P^{(w_k)}_{V-A}(Q^2)$ using ALEPH experimental data \cite{ALEPH13} 
for the spectral functions $\r_V(s)$ and $\br_A(s)$ for 
$s\le s_0=s_{\rm switch}$,\footnote{For details on the handling of the
ALEPH data, see Ref.~\cite{alphas14}.} and
approximating the difference $\r_V(s)-\br_A(s)$ by Eq.~(\ref{approx}) for
$s\ge s_0=s_{\rm switch}$, with values for the DV parameters from the combined
$V$ and $A$ channel fits of Ref.~\cite{alphas14}. We will choose
$s_{\rm switch}$ to be the upper end of an ALEPH bin, obtaining, in the notation
adopted in Ref.~\cite{alphas14},
\begin{eqnarray}
\label{PiALEPH}
\bP_{V-A}^{(w)}(Q^2)&=&\sum_{{\tt sbin}<s_{\rm switch}}
\int_{{\tt sbin}-{\tt dsbin}/2}^{{\tt sbin}+{\tt dsbin}/2}
\!\!\!\!\!\!\!\!\!\!\!ds\;w\left(\frac{s}{s_{\rm switch}}\right)\;
\frac{\r_V({\tt sbin})-\br_A({\tt sbin})}{{\tt sbin}+Q^2}\nonumber\\
&&+\int_{s_{\rm switch}}^\infty ds\;w\left(\frac{s}{s_{\rm switch}}\right)\;\frac{\r^{\rm DV}_V(s)-\r^{\rm DV}_A(s)}{s+Q^2}\ .
\end{eqnarray}
Here we will choose $s_{\rm switch}=1.55$~GeV$^2$,
the value of $s_{\rm min}$ which produced
the best fit to the weighted spectral integrals in our extraction of 
$\a_s$ in Ref.~\cite{alphas14}.
Since we are only interested in the
behavior of $\bP_{V-A}^{(w)}(Q^2)$ at values of $Q^2\ll 1.55$~GeV$^2$,
the right-hand side of Eq.~(\ref{PiALEPH}) is very insensitive to the precise choice
of $s_{\rm switch}$, and varying it within the range of $s_{\rm min}$ values for
which we obtained good fits in Ref.~\cite{alphas14} has no effect on either the
values or errors we will obtain for the LECs below.   

We have fully propagated all errors and correlations in the results 
we will report on below.
In particular, the DV parameter values used in Eq.~(\ref{PiALEPH}) are
correlated with the data, and we have computed these correlations using the
linear error propagation method summarized in the appendix of Ref.~\cite{US2}
(see, in particular, Eq.~(A.4) of that reference, which can be used to express
the parameter-data covariances in terms of the data covariance matrix).

For the $us$ data needed in order to evaluate the 
strange spectral integrals in Eq.~(\ref{vpmaimsralt})
we will use exactly the same treatment and input as used 
in Ref.~\cite{bgjmp13}, and we refer to Sec.~III.C of that article for details.
We collect the values of all other external input parameters:
\begin{eqnarray}
\label{inputpar}
m_\p&=&139.57~\mbox{MeV}\ ,\\
m_K&=& 495.65~\mbox{MeV}\ ,\nonumber\\
m_\h&=& 547.85~\mbox{MeV}\ ,\nonumber\\
f_\p&=& 92.21(14)~\mbox{MeV}\ ,\nonumber\\
f_K&=& 110.5(5)~\mbox{MeV}\ ,\nonumber\\
L_5^r(\m=770~\mbox{MeV})&=&0.84(38)\times 10^{-3}
\ ,\qquad \mbox{\cite{MILC9A}}\ ,\nonumber\\
L_9^r(\m=770~\mbox{MeV})&=&5.93(43)\times 10^{-3}\ ,\qquad \mbox{\cite{bt02}}
\ .\nonumber
\end{eqnarray}
For the value of $L_5^r$ we choose the value reported of Ref.~\cite{MILC9A}, 
which has a larger error than the value of Ref.~\cite{bj11} we used in 
Ref.~\cite{bgjmp13}. The comparison with other values in Ref.~\cite{FLAG} 
shows that the error quoted in Eq.~(\ref{inputpar}) above covers these 
other values, and we thus consider its use to represent a conservative choice.

\section{\label{results} Results }
In this section, we present the results of our analysis, dividing the 
presentation into several parts. We first present our values for 
$L_{10}^{\rm eff}$ and $C_{87}^{\rm eff}$, which are based purely 
on non-strange $\t$-decay data, then derive additional constraints 
employing also the lattice data, and finally use the $us$ spectral 
data to obtain further constraints via the flavor-breaking IMFESRs.
The output is a determination of 
$L_{10}^r$, $C_{61}^r$, $C_{87}^r$, $C_{12}^r-C_{61}^r+C_{80}^r$ and
$C_{13}-C_{62}+C_{81}$, in terms
of the known values for $L_5^r$ and $L_9^r$. In the final subsection, we
switch gears, and consider what we can learn from the ALEPH data about the
OPE coefficients $C_{6,V-A}$ and $C_{8,V-A}$.

\subsection{\label{effective} Effective LECs}
We first give the values of $L_{10}^{\rm eff}$ and $C_{87}^{\rm eff}$,
which follow directly from Eq.~(\ref{alt}) and the evaluation of Eq.~(\ref{PiALEPH}). For the three
different weights we find
\begin{subequations}
\label{L10eff}
\begin{eqnarray}
L_{10}^{\rm eff}&=&-6.482(64)\times 10^{-3}\ ,\qquad 
(\mbox{from}\ \bP_{V-A}(0))\ ,
\label{L10effa}\\
&=&-6.486(64)\times 10^{-3}\ ,\qquad (\mbox{from}\ \bP^{(w_1)}_{V-A}(0))\ ,
\label{L10effb}\\
&=&-6.446(50)\times 10^{-3}\ ,\qquad (\mbox{from}\ \bP^{(w_2)}_{V-A}(0))\ .
\label{L10effc}
\end{eqnarray}
\end{subequations}
These values are consistent with those found
employing OPAL data in Ref.~\cite{bgjmp13}, but more precise, with errors about a 
factor two smaller. The contribution from the DV part of the spectral function
in Eq.~(\ref{PiALEPH}) ranges from 3\% for the first estimate in Eq.~(\ref{L10eff}) to 
about 1\% for the third estimate. Their size is thus comparable with the 
quoted errors, suggesting that the uncertainty in the DV part due to the
use of a model for DVs, Eq.~(\ref{ansatz}), is negligible.
From the slope of $\bP_{V-A}(Q^2)$ at zero we find
\begin{equation}
\label{C87eff}
C_{87}^{\rm eff}=8.38(18)\times 10^{-3}\ \mbox{GeV}^2\ .
\end{equation}
In this case, the contribution from the DV part is about 1\%. 
These results are in good agreement with those in Refs.~\cite{GPP3,Dominguezetal}.
However, our errors are somewhat smaller, and our analysis
employs versions of the DV contributions fitted individually
in the $V$ and $A$ channels, as required by the data, 
avoiding the simplifications employed in
Ref.~\cite{GPP3,Dominguezetal} (see also the discussion in
Ref.~\cite{bgjmp13}).

{}From our best value for $L_{10}^{\rm eff}$, Eq.~(\ref{L10effc}), and using 
Eq.~(\ref{chptconna}), we find the constraint
\begin{eqnarray}
\label{VmAconstraint}
L_{10}^r&=&0.6576L_{10}^{\rm eff}+0.001161-0.1712L_9^r+0.08220({\cal C}_0^r
+{\cal C}_1^r)\\
&=&-0.004094(33)_{\rm ALEPH}(74)_{L_9^r}+0.08220({\cal C}_0^r
+{\cal C}_1^r)
\ .\nonumber
\end{eqnarray}
Together with information from the lattice on the combinations ${\cal C}_0^r$ 
and ${\cal C}_1^r$, this constraint will yield an estimate for $L_{10}^r$.
Likewise, Eq.~(\ref{chptconnb}) leads to the constraint \cite{bgjmp13}
\begin{equation}
\label{C87effC87}
C_{87}^{\rm eff}=C_{87}^r+0.292L_9^r+0.00155\ \mbox{GeV}^{-2}\ ,
\end{equation}
and, with Eqs.~(\ref{C87eff}) and~(\ref{inputpar}), this leads to the estimate
\begin{equation}
\label{C87res}
C_{87}^r=5.10(22)\times 10^{-3}\ \mbox{GeV}^{-2}\ .
\end{equation}

\begin{boldmath}
\subsection{\label{lattice} Constraints using the lattice}
\end{boldmath}
As shown in Ref.~\cite{Boyleetal14}, and noted already above,
useful independent constraints on $L_{10}^r$, ${\cal C}_0^r$ and 
${\cal C}_1^r$ can be obtained by considering the difference of 
the lattice-ensemble, $E$, and physical-mass, continuum results 
for $\overline{\Pi}_{V-A}(Q^2)$, evaluated at the same $Q^2$. 
In what follows, we will use the superscript $phys$ to specify
quantities evaluated with physical values of the masses and decay
constants. 

Recasting the NNLO representation, Eq.~(\ref{deltadeltapibar}), 
in the form
\begin{equation}
\d_{10}^{\rm L,E}\, L_{10}^r\, +\, \delta^{L,E}_0 {\cal C}_0^r\,
+\, \delta^{\rm L,E}_1 {\cal C}_1^r\, =\, \D\bP_{V-A}^{\rm L,E}(Q^2)
-\, \Delta R_{\pi K}^{\rm L,E}(\mu ,Q^2; L_9^r)\, \equiv\,
T^{\rm L,E}(Q^2)\ ,
\label{deltadeltapibarnnloalt}\end{equation}
one notes that the left-hand side is explicitly $Q^2$-independent, while the
right-hand side is the difference of two $Q^2$-dependent terms. For 
values of $Q^2$ for which the NNLO representation 
is reliable, the $T^{\rm L,E}(Q^2)$ for fixed ensemble $E$ but different 
$Q^2$ should be compatible within errors, thus providing non-trivial 
self-consistency checks. We will denote 
the average of the $T^{\rm L,E}(Q^2)$ for an ensemble satisfying these 
self-consistency tests by $T^{\rm L,E}$.

The explicit expression for 
$\Delta R_{\pi K}^{\rm L,E}(\mu ,Q^2; L_9^r)\equiv
R_{\pi K}^{\rm L,E}(\mu ,Q^2;L_9^r) \, -\,  R_{\pi K}^{phys}(\mu ,Q^2;L_9^r)$
 follows from the results 
of Ref.~\cite{ABT}, but is very lengthy and hence not given here. The 
explicit expressions for the mass-dependent constants appearing on 
the LHS of Eq.~(\ref{deltadeltapibarnnloalt}) are
\begin{eqnarray}
\d_{10}^{\rm L,E}&=&
32\, \left[ \left( 2\mu_\pi +\mu_K\right)^{\rm L,E} \, -\, 
\left( 2\mu_\pi +\mu_K\right)^{phys}\right]\ , \label{lattconstnnloleccoefs}\\
\delta^{\rm L,E}_0&\equiv& {\frac{[m_\pi^2]^{\rm L,E}}{[m_\pi^2]^{phys}}}
\, -\, 1\ , \nonumber\\
\delta^{\rm L,E}_1&\equiv& {\frac{\left[ m_\pi^2+2m_K^2\right]^{\rm L,E}}
{\left[ m_\pi^2+2m_K^2\right]^{phys}}}\, -\, 1\ .
\nonumber\end{eqnarray}

The lattice data we employ in forming the lattice-minus-continuum
constraints are obtained using the $n_f=2+1$ domain wall fermion 
ensembles of the RBC/UKQCD collaboration. Details of the underlying 
simulations may be found in Refs.~\cite{rbcukqcdfine11,rbcukqcdcoarse12}, 
with updated values of the lattice spacings $a$, obtained after 
incorporating results from the new physical point ensembles, given 
in Ref.~\cite{rbcukqcdphyspt14}. 

We have used the following criteria 
in deciding on the choice of ensembles and $Q^2$ values to be employed.
First, we restrict our attention to ensembles with $m_\pi < 350$ MeV. 
Second, we require the ensemble to have sufficiently many $Q^2$ 
points in the expected range of validity of the representation, 
Eq.~(\ref{deltadeltapibar}), that meaningful self-consistency tests 
can be performed. With the errors at the lowest $Q^2$ point turning
out to be very large for all ensembles considered, this means that
a minimum of two additional such $Q^2$, or three in total, are required. 
Finally, we identify the range of validity of the representation 
Eq.~(\ref{deltadeltapibar}) as follows. We first note that the 
supplemented NNLO representation of $\overline{\Pi}_{V-A}(Q^2)$, 
discussed in more detail in Refs.~\cite{bgjmp13,gmp14}, and the 
corresponding NNLO representation given above, both yield the same 
representation of the lattice-continuum difference 
$\D\bP_{V-A}^{\rm L,E}(Q^2)$. The supplemented form of 
$\overline{\Pi}_{V-A}(Q^2)$ incorporates resonance-induced NNNLO 
contributions analogous to those already encoded in the NNLO contribution 
proportional to $C_{87}^r$ through the inclusion of an additional analytic 
term $CQ^4$. The inclusion of such a term was shown to extend the 
reliability of the supplemented version of the representations to 
significantly larger $Q^2$ in both the $V-A$ and $V$ correlator 
cases~\cite{bgjmp13,gmp14}. Here we investigate the supplemented
NNLO fit by fixing $L_{10}^{\rm eff}$ and $C_{87}^{\rm eff}$ to the values 
given in Eqs.~(\ref{L10eff}) and~(\ref{C87eff}), and fitting the additional effective mass-independent NNNLO 
LEC, $C$, to the dispersive results for $\overline{\Pi}_{V-A}(Q^2)$ 
in the window $0<Q^2<Q^2_{\rm max}$. The range of validity of the supplemented
form is then identified as the largest $Q^2_{\rm max}$ for which such a fit is 
successful within errors. The results of this exploration show that the 
supplemented form can be employed up to $\sim 0.25\ {\rm GeV}^2$, but not 
beyond. We thus restrict our attention further to ensembles having at least 
three $Q^2$ in the region below $0.25\ {\rm GeV}^2$.

Three RBC/UKQCD ensembles satisfy the above criteria,
two coarse ensembles with $1/a=1.379(7)$ GeV and $m_\pi = 172$ and 
$250$ MeV, and one intermediate ensemble with $1/a=1.785(5)$ GeV 
and $m_\p =340$ MeV \cite{rbcukqcdphyspt14}. 
These are labelled $E=1$, $2$ and $3$ in
what follows. The coarse ensembles have eight points each below
$0.25\ {\rm GeV}^2$, the intermediate ensemble has four. The results for
the $V-A$ correlators for all three of these ensembles pass the 
self-consistency tests discussed above.

The lattice-continuum constraints for the first two cases were obtained 
previously in Ref.~\cite{Boyleetal14}. While the results of 
Ref.~\cite{rbcukqcdphyspt14} lead to a small shift in the value
of $1/a$ for these ensembles, this change affects the constraints
only through the values of the $Q^2$ at which the dispersive 
representation of the physical-mass correlator is evaluated,
the values of $a^2Q^2$ being fixed by the lattice size. The small 
resulting $Q^2$ changes turn out to have no impact 
on the resulting $T^{\rm L,E}$ averages for these ensembles to 
the number of significant figures previously quoted. The 
resulting average $T^{\rm L,E}$ are thus those given in Ref.\cite{Boyleetal14},
\begin{eqnarray}
T^{\rm L,1}&&\, =\, 0.0007(17)\ ,\nonumber\\
T^{\rm L,2}&&\, =\, 0.0039(21)\ .
\label{Taveensemble1and2}\end{eqnarray}

The third ensemble has a relative error on $af_\pi$
a factor of $2$ smaller than that for the two coarse ensembles, and
hence a smaller uncertainty in the pion-pole subtraction involved
in forming $\overline{\Pi}^{\rm L,E}_{V-A}(Q^2)$ from the directly measured
unsubtracted version. In order to improve further the associated constraint, 
the statistics for this ensemble were increased using 
multiple time sources. This increase in statistics 
was greatly aided by the use of the HDCG algorithm of Ref.~\cite{Boyle:2014rwa},
employed in performing the propagator inversions.
The covariances of the corresponding lattice-minus-continuum
differences for different $Q^2$ (equal to the sum of the covariances of 
the corresponding lattice and dispersive results) are strongly dominated by 
the lattice contributions. With the covariance matrix available, 
the average constraint value, $T^{\rm L,3}$, can be obtained by a standard,
fully correlated $\chi^2$ fit to the four $T^{\rm L,3}(Q^2)$ with
$Q^2<0.25\ {\rm GeV}^2$ available for this ensemble. The result is
\begin{equation}
T^{\rm L,3}\, =\, 0.00625(48)\ .
\label{Taveensemble3}\end{equation}
The error here reflects only the errors on (and correlations among) 
the lattice results at different $Q^2$ and dispersive results at 
different $Q^2$. Additional errors due to the 
uncertainties on the input quantities $L_5^r(\mu )$ and $L_9^r(\mu )$, 
which are $100\%$ correlated with the analogous uncertainties entering the
$\overline{\Pi}_{V-A}(0)$ and flavor-breaking IMFESR constraints, are handled
separately below.

\vskip0.7cm
\begin{boldmath}
\subsection{\label{IMFESR} Constraints using IMFESRs}
\end{boldmath}
In this subsection, we evaluate $\D\bP_T(0)$ with $T=V$, and $V\pm A$
from Eq.~(\ref{vpmaimsralt}), for values of $s_0$ between $2.15~\mbox{GeV}^2$
and $m_\t^2$.   Of course, $\D\bP_T(0)$ is independent of $s_0$, implying that
evaluating the right-hand side of the expressions in Eq.~(\ref{vpmaimsralt}) and
checking for $s_0$ independence provides a self-consistency check on the
use of the OPE, and the assumption that DVs can be neglected. The values
we find are, using $w(y)=w_{\rm DK}(y)$ in Eq.~(\ref{vpmaimsralt}),
\begin{eqnarray}
\label{DeltaPivalues}
\D\P_{V}(0)&=&0.0224(9)
\ ,\\
\D\bP_{A}(0)&=&0.0113(8)
\ ,\nonumber\\
\D\bP_{V+A}(0)&=&0.0338(10)
\ ,\nonumber\\
\D\bP_{V-A}(0)&=&0.0111(11)
\ .\nonumber
\end{eqnarray}
The result for $\D\bP_A(0)$ obtained from the analogous $A$ channel IMFESR 
is also included for completeness. While the axial case is not independent 
of the others, performing the axial analysis directly is the most 
straightforward way to take into account the correlations amongst 
the other channels and arrive at the correct error for the $A$ case.
The quoted errors take into account the experimental errors in the ALEPH data,
the uncertainties in the estimates of the OPE contribution, and the 
small residual $s_0$-dependence observed as $s_0$ is varied over the 
analysis window noted above.  As a further check of the self-consistency 
of the values in Eq.~(\ref{DeltaPivalues}), 
we have rerun the analysis using $w(y)=\hat{w}(y)$ in place of $w_{\rm DK}(y)$, 
and find results compatible with those obtained using $w_{\rm DK}(y)$ to well 
within the errors quoted in Eq.~(\ref{DeltaPivalues}). 
The analysis method leading to the values~(\ref{DeltaPivalues}) is 
identical to that of Ref.~\cite{bgjmp13}, to which we refer for a detailed 
discussion.

From the value for $\D\bP_{V+A}(0)$ and Eq.~(\ref{nlonnlocontributions}) we
find, using the values of $L_5^r$ and $L_9^r$ in Eq.~(\ref{inputpar}),
\begin{eqnarray}
\label{VpA}
C_{12}^r+C_{61}^r+C_{80}^r&=&0.00237(13)_{\D\bP}(4)_{L_5^r}(8)_{L_9^r}\ 
\mbox{GeV}^{-2}\\
&=&0.00237(16)\ \mbox{GeV}^{-2}
\ ,\nonumber
\end{eqnarray}
where the subscript $\D\bP$ refers to the error coming from the value for
$\D\bP_{V+A}(0)$ in Eq.~(\ref{DeltaPivalues}), and on the second line we have
combined errors in quadrature. The sum rules for $\D\bP_{V-A}(0)$ and
$\D\P_V(0)$ provide two independent constraints 
involving combinations of $L_{10}^r$ and other NNLO LECs:
\begin{eqnarray}
\label{VmA}
2.125L_{10}^r-32(m_K^2-m_\p^2)(C_{12}^r-C_{61}^r+C_{80}^r)&=&
-0.0034(11)_{\D\bP}(3)_{L_5^r}(6)_{L_9^r}\ \mbox{GeV}^{-2}\nonumber\\
&=&
-0.0034(13)\ \mbox{GeV}^{-2}\ ,
\end{eqnarray}
and
\begin{eqnarray}
\label{V}
1.062L_{10}^r+32(m_K^2-m_\p^2)C_{61}^r&=&
0.0071(9)_{\D\bP}(3)_{L_5^r}(6)_{L_9^r}\ \mbox{GeV}^{-2}\\
&=&
0.0068(11)\ \mbox{GeV}^{-2}\ .\nonumber
\end{eqnarray}

Employing the constraints Eqs.~(\ref{VmAconstraint}),
~(\ref{VmA}) and the $E=1,2,3$ versions of Eq.~(\ref{deltadeltapibarnnloalt}), 
with Eqs.~(\ref{Taveensemble1and2}) and ~(\ref{Taveensemble3}) 
as input on the right-hand side and Eq.~(\ref{lattconstnnloleccoefs}) as input on
the left-hand side, we find
\begin{eqnarray}
\label{LEClattice}
L_{10}^r&=&-0.00350(11)(13)_{L_5,L_9}=-0.00350(17) \ ,\\
{\cal C}_0^r&=&-0.00035(9)(4)_{L_5,L_9}=-0.00035(10) \ ,\nonumber\\
{\cal C}_1^r&=&0.0075(13)(8)_{L_5,L_9}=0.0075(15) \ .\nonumber
\end{eqnarray}
Here the first error quoted comes primarily from the errors
on the lattice data,\footnote{Lattice errors in the 
differences~(\ref{DeltaPivalues}) dominate over the errors coming 
from the $\t$ spectral data.} while the second error comes 
from the errors in $L_5^r$ and $L_9^r$.
The value of ${\cal C}_0^r$ translates directly into
\begin{equation}
\label{C12mC61pC80}
C_{12}^r-C_{61}^r+C_{80}^r=-0.00056(15)\ \mbox{GeV}^{-2}\ ,
\end{equation}
the value of ${\cal C}_1^r$ into
\begin{equation}
\label{C13mC62pC81}
C_{13}^r-C_{62}^r+C_{81}^r=0.00046(9)\ \mbox{GeV}^{-2}\ ,
\end{equation}
while the value of $L_{10}^r$ in Eq.~(\ref{LEClattice}) together with Eq.~(\ref{V})
leads to the estimate
\begin{equation}
\label{C61}
C_{61}^r=0.00146(15)\ \mbox{GeV}^{-2}\ .
\end{equation}
For completeness, the result for the NNLO LEC
combination entering the flavor-breaking $A$ IMFESR is
\begin{equation}
\label{C12pC80}
C_{12}^r+C_{80}^r=0.00090(9)\ \mbox{GeV}^{-2}\ .
\end{equation}

\vskip0.7cm
\begin{boldmath}
\subsection{\label{condensates} $V-A$ condensates}
\end{boldmath}
In addition to the LECs extracted in the previous subsections, $\bP_{V-A}(Q^2)$
also provides information on the OPE coefficients $C_{6,V-A}$ and $C_{8,V-A}$.
Adapting Eq.~(4.20) of Ref.~\cite{bgjmp13} to the case of the ALEPH data, following
the notation of Ref.~\cite{alphas14}, used also in Eq.~(\ref{PiALEPH}) above, these two
coefficients are given by 
\begin{subequations}
\label{C68}
\begin{eqnarray}
C_{6,V-A}&=&\sum_{{\tt sbin}<s_{\rm switch}}\int_{{\tt sbin}-{\tt dsbin}/2}^{{\tt sbin}+{\tt dsbin}/2}\!\!\!\!\!\!\!\!\!\!\!ds\;(s-s_{\rm switch})^2\left(\r_V({\tt sbin})-\br_A({\tt sbin})\right)
\label{C68a}\\
&&-2f_\p^2\left(m_\p^2-s_{\rm switch}\right)^2+\int_{s_{\rm switch}}^\infty ds\;(s-s_{\rm switch})^2\left(\r_V^{\rm DV}(s)- \r_A^{\rm DV}(s)\right)\ ,\nonumber\\
C_{8,V-A}&=&-\sum_{{\tt sbin}<s_{\rm switch}}\int_{{\tt sbin}-{\tt dsbin}/2}^{{\tt sbin}+{\tt dsbin}/2}\!\!\!\!\!\!\!\!\!\!\!ds\;(s-s_{\rm switch})^2(s+2s_{\rm switch})\left(\r_V({\tt sbin})-\br_A({\tt sbin})\right)\nonumber
\\
&&+2f_\p^2 \left(m_\p^2-s_{\rm switch}\right)^2(m_\p^2+2s_{\rm switch})
\nonumber\\
&&-\int_{s_{\rm switch}}^\infty ds\;(s-s_{\rm switch})^2(s+2s_{\rm switch})\left(\r_V^{\rm DV}(s)-\r_A^{\rm DV}(s)\right)\ .\label{C68b}
\end{eqnarray}
\end{subequations}
The first of these two expressions 
involves contributions proportional to $C_{2k,V-A}$ for $k=1,\ 2,\ 3$, 
but the leading-order expressions~\cite{FNR,NLOC4}
\begin{subequations}
\label{C2C4}
\begin{eqnarray}
\hspace{-0.2cm}C_{2,V-A}
&\!\!\!=&\!\!\!-\frac{\a_s(\m^2)}{\p^3}\,m_u(\m^2)m_d(\m^2)
\left(1\!-\!\frac{\a_s(\m^2)}{\p}\left(\frac{17}{4}
\log{\frac{Q^2}{\m^2}}+c\right)\!\right)+\dots,\label{C2C4a}\\
\hspace{-0.2cm}C_{4,V-A}&\!\!\!=&\!\!\!-\frac{8}{3}\frac{\a_s}{\p}
f_\p^2 m_\p^2+\dots\ ,\label{C2C4b}
\end{eqnarray}
\end{subequations}
with $c$ a numerical constant whose value is not required in what follows,
suggest that the $D=2$ and $D=4$ terms
are numerically negligible.   A similar observation holds for the second
of these expressions.
The use of Eq.~(\ref{C68}) to estimate $C_{6,V-A}$ and $C_{8,V-A}$ 
thus rests on the assumption that these estimates for the $D=2$ and $D=4$ 
contributions are sufficiently reliable, which is equivalent to assuming 
that the two Weinberg sum rules hold exactly.
In this case, Eq.~(\ref{C68}) leads to the estimates
\begin{eqnarray}
\label{C6C8num}
\mbox{ALEPH}:\qquad C_{6,V-A}&=&(-3.16\pm 0.91)\times 10^{-3}\ \mbox{GeV}^6\ ,
\\
C_{8,V-A}&=&(-13.0\pm 5.5)\times 10^{-3}\ \mbox{GeV}^8\ .
\nonumber
\end{eqnarray}
These results correspond to the choice
$s_{\rm switch}=2.2\ \mbox{GeV}^2$, which yields the smallest 
estimate for the errors on $C_{6,V-A}$ and $C_{8,V-A}$. We have checked 
that the central values remain stable as a function of $s_{\rm switch}$ 
as $s_{\rm switch}$ is varied between $s_{\rm min}=1.55\ {\rm GeV}^2$ and 
$m_\t^2$. The results of Eq.~(\ref{C6C8num}) are in agreement 
within errors with those of Refs.~\cite{GPP3,Dominguezetal}.

Let us compare these values with those we found from the OPAL data in
Ref.~\cite{bgjmp13}:
\begin{eqnarray}
\label{C6C8t3}
\mbox{OPAL}:\qquad C_{6,V-A}&=&(-6.6\pm 1.1)\times 10^{-3}\ \mbox{GeV}^6
\ ,\\
C_{8,V-A}&=&(5\pm 5)\times 10^{-3}\ \mbox{GeV}^8
\ .\nonumber
\end{eqnarray}
These values differ by $2.4\ \s$ from the ALEPH values 
in Eq.~(\ref{C6C8num}).

Instead of the weights employed
in Eq.~(\ref{C68}), one can use the weight $s^2$ to estimate $C_{6,V-A}$, and
the weight $s^3$ to estimate $C_{8,V-A}$. If we do so, we find the values
\begin{eqnarray}
\label{C6C8direct}
C_{6,V-A}&=&(-4.4\pm 5.2)\times 10^{-3}\ \mbox{GeV}^6\ ,\\
C_{8,V-A}&=&(-8\pm 18)\times 10^{-3}\ \mbox{GeV}^8\ .
\nonumber
\end{eqnarray}
This suggests that central values of $C_{2,V-A}$ and $C_{4,V-A}$ 
produced by the ALEPH data are larger than those following from Eq.~(\ref{C2C4}).
Direct determination of these coefficients from the data, using the
weights $1$ and $s$ on the right-hand side of Eq.~(\ref{C68}), 
yield values $C_{2,V-A}=(0.14\pm 0.24)\times 10^{-3}\ \mbox{GeV}^2$ and
$C_{4,V-A}=(0.1\pm 1.2)\times 10^{-3}\ \mbox{GeV}^4$. These central values
are indeed larger and explain
the difference between the estimates~(\ref{C68}) and~(\ref{C6C8direct}),
though the results are consistent with Eq.~(\ref{C2C4}) within 
errors.
The smaller errors reported in Eq.~(\ref{C6C8t3}) are thus a consequence of the
{\it assumption} that $C_{2,V-A}$ and $C_{4,V-A}$ can be neglected, or, 
equivalently, the assumption that the first and second Weinberg sum rules 
are exactly satisfied. 

In view of the discussion above, we conclude that current data 
lead to somewhat conflicting estimates for
$C_{6,V-A}$ and $C_{8,V-A}$.
This is not surprising, because these coefficients are sensitive to the 
large-$s$ region of the data. In addition, we observe
the contribution from the DV terms to the expressions on the right-hand 
side of Eq.~(\ref{C68}), \ie, the size of the DV contributions which must be 
subtracted to arrive at the values reported in Eq.~(\ref{C6C8num}), are 
non-negligible: about $-0.28\times 10^{-3}\ \mbox{GeV}^6$ and 
$3\times 10^{-3}\ \mbox{GeV}^8$ 
for $C_{6,V-A}$ and $C_{8,V-A}$, respectively.

\vskip0.7cm
\section{\label{conclusion} Conclusion}
Using the recently updated and corrected ALEPH results for the 
non-strange $V$ and $A$ hadronic $\tau$ decay distributions,
existing results for the corresponding strange decay distributions,
and updated lattice results for the non-strange $V-A$ polarization
at heavier than physical meson masses, we have
produced improved determinations of the NLO chiral LEC $L_{10}^r$
and a number of NNLO LEC combinations. Those results are given
in Eqs.~(\ref{VpA}) and (\ref{LEClattice}) to~(\ref{C12pC80}). We have also used the 
non-strange ALEPH data to extract the dimension six and eight
condensates appearing in the OPE representation of the $V-A$
polarization. These results are given in Eq.~(\ref{C6C8num}).

The improvements produced by using the new lattice results and 
the new ALEPH data in place of the old OPAL data
reduce the fit component of the errors on $L_{10}^r$ and the
$1/N_c$-suppressed NNLO LEC combination $C_{13}^r-C_{62}^r+C_{81}^r$
by a factor of roughly $2.5$ compared to our earlier analysis. The 
fit errors on the remaining NNLO LECs are about $2/3$ of those of 
the previous analysis. Taking, as in Ref.~\cite{bgjmp13}, $25\%$ as an
estimate of the expected reduction in size of contributions in going from
one order to the next in the chiral expansion, we would expect
the uncertainties from the neglect of NNNLO and higher contributions
to be roughly $6\%$ for $L_{10}^r$ and $25\%$ for the NNLO LECs.
With the current fit errors, the total errors on all the LECs
determined have reached these levels, suggesting that the optimal 
practical precision one can expect for an NNLO analysis
has now been attained.

We find, in contrast, that using the higher precision ALEPH
data produces essentially no improvement in the accuracy
of the determination of the dimension six and eight $V-A$ OPE
condensates. Our results for these quantities are in agreement
with those of other ALEPH-based analyses, and show about $2.4\sigma$ 
discrepancies with the corresponding results obtained using
the OPAL data. These discrepancies presumably reflect additional
systematic uncertainties encountered in attempting to extract
these small higher dimension contributions from existing data,
and should be kept in mind if results based on one or the other
of the two data sets are employed in other contexts.

\vspace{3ex}
\noindent {\bf Acknowledgments}
\vspace{3ex}

DB thanks the  Department of Physics of the
Universitat Aut\`onoma de Barcelona, and 
KM and SP thank the Department of Physics and Astronomy at San 
Francisco State University for hospitality. DB is supported by the 
S\~ao Paulo Research Foundation (Fapesp) grant 14/50683-0;
MG is supported in part by the US Department of Energy, AF, RH, RL and KM are
supported by grants from the Natural Sciences and Engineering Research 
Council of Canada, and SP is supported by CICYTFEDER-FPA2011-25948, 
2014~SGR~1450, the Spanish Consolider-Ingenio 2010 Program
CPAN (CSD2007-00042). Propagator inversions for the improved
lattice data were performed on the STFC funded ``DiRAC'' BG/Q
system in the Advanced Computing Facility at the University of
Edinburgh.


\end{document}